\documentclass[aps,pre,floatfix,superscriptaddress,twocolumn,showpacs]{revtex4}

\usepackage{epsfig}
\usepackage{amsmath}
\usepackage{graphicx,psfrag}
\usepackage{dcolumn}
\usepackage{bm}
\usepackage{slashed}
\usepackage{color}

\newcommand{\beq}{\begin{equation}}
\newcommand{\eeq}{\end{equation}}
\newcommand{\bea}{\begin{eqnarray}}
\newcommand{\eea}{\end{eqnarray}}

\newcommand{\nn}{\nonumber\\}

\newcommand\fig[1]     {Fig.\,{\ref{#1}}}

\def\eq#1{(\ref{#1})}
\def\s0#1#2{\mbox{\small{$ \frac{#1}{#2} $}}}
\def\0#1#2{\frac{#1}{#2}}

\def\mr#1{{\mathrm{#1}}}

\begin{document}

\title{Improved efficiency of heat generation in nonlinear dynamics of magnetic nanoparticles} 

\author{J. R\'acz} 
\affiliation{University of Debrecen, H-4010 Debrecen P.O.Box 105, Hungary}
\affiliation{Institute of Nuclear Research, P.O.Box 51, H-4001 Debrecen, Hungary} 

\author{P. F. de Ch\^atel}
\affiliation{MTA-DE Particle Physics Research Group, H-4010 Debrecen P.O.Box 105, Hungary}

\author{I. A. Szab\'o}
\affiliation{University of Debrecen, H-4010 Debrecen P.O.Box 105, Hungary}

\author{L. Szunyogh}
\affiliation{Department of Theoretical Physics and MTA-BME Condensed Matter Research Group
Budapest University of Technology and Economics, H-1111 Budapest Budafoki 8., Hungary}

\author{I. N\'andori}
\affiliation{MTA-DE Particle Physics Research Group, H-4010 Debrecen P.O.Box 105, Hungary}
\affiliation{Institute of Nuclear Research, P.O.Box 51, H-4001 Debrecen, Hungary} 

\begin{abstract} 
The deterministic Landau-Lifshitz-Gilbert equation has been used to investigate 
the nonlinear dynamics of magnetization and the specific loss power in magnetic 
nanoparticles with uniaxial anisotropy driven by a rotating magnetic field. We propose a 
new type of applied field, which is "simultaneously rotating and alternating", i.e. the 
direction of the rotating external field changes periodically. We show that a more efficient 
heat generation by magnetic nanoparticles is possible with this new type of applied field
and we suggest its possible experimental realization in cancer therapy which requires the 
enhancement of loss energies. 
\end{abstract}

\pacs{75.75.Jn, 82.70.-y, 87.50.-a, 87.85.Rs}
\maketitle

\section{Introduction}
\label{sec_intro}
Nonlinear dynamics of the magnetization in single-domain ferromagnetic nanoparticle systems has received 
a considerable attention due to a wide range of their applicability such as magnetic resonance imaging, 
ultrahigh density magnetic data recording, spintronics, ferrofluids and in biomedical engineering, in particular, 
drug delivery or hyperthermia \cite{ferrofluid,ferrohydro,biomedical}. Among many forms of hyperthermia, the 
local induction of heat via magnetic nanoparticles seems a fruitful strategy with promising preclinical results in 
different cancer modes \cite{Stauffer, Johannsen}. The unique feature of magnetic nanoparticle hyperthermia 
is that the energy is transported in the body by means of an ac magnetic field. At present the clinical application 
is partly limited due to the efficacy of the heat transfer and the poor controllability of temperature parameters 
\cite{Bellizzi}. Thus, the study of relaxation mechanisms of magnetic nanoparticles is a very active research field, 
both in its theoretical and material-science aspects.

Among many controllable parameters, the applied external magnetic field is one of the most easily variable to 
increase the efficiency of heat generation. Indeed, there is an increasing interest in the literature to consider the 
case of rotating external magnetic field  
\cite{Bertotti,Denisov2006,Chatel,Cantillon,Ahsen2010,Raikher,Denisov_thermal,Nandori,Lyutyy,Lyutyy_energy,Chen} 
instead of the commonly applied alternating one, see for example \cite{stochastic_llg_lin}. Some of these studies 
compare the efficiency of the two types of applied fields. For example, in \cite{Chen} it was argued that the use of 
an alternating-like external field is favourable, in particular an orthogonal synchronised bi-directional field is proposed 
as a most efficient heat generation. Another experimental work \cite{Ahsen2010} suggests that the alternating and 
circulating applied fields produce the same heating efficiency in the limit of low-frequency. 

In order to clarify the above question and to look for the most efficient heat generation one has to take into 
account two other important effects: the role of magnetic anisotropy which is inevitably present in magnetic nanoparticles 
and the influence of thermal fluctuations. For example, in \cite{Chatel}, isotropic nanoparticles were considered 
without taking into account the thermal effects and the alternating applied field was found to be considerably 
more favourable than the rotating one. However, it was shown in \cite{Raikher} that the effect of thermal 
fluctuation modifies the results obtained for the alternating field with isotropic nanoparticles and results in a 
small difference between the heating efficiency of the rotating and alternating applied fields in the limit of small 
frequency. Let us note, the inclusion of thermal fluctuations \cite{Raikher} only slightly modifies the findings of 
the rotating field obtained for isotropic nanoparticles without thermal effects \cite{Chatel}.

Magnetic anisotropy can also influence this picture, specially in case of the rotating applied field. An important 
feature of the rotating external field is the presence of stable steady states (precession modes). Independently 
of the initial positions, the magnetic moments of nanoparticles tend to the steady state and the dissipated 
energy can be easily calculated. However, for relatively large anisotropy, more than one steady states appear 
\cite{Bertotti}. Thus, one can study the influence of the transition between these modes on the heating 
\cite{Denisov2006,Denisov_thermal}. For example, in Refs. \cite{Lyutyy,Lyutyy_energy} the effect of strong 
anisotropy on the heating efficiency has been studied and its enhancement is shown for large frequencies 
near the boundary of various regimes of forced precession (steady states) of the parameter space in case of 
a rotating applied field. Although, it was argued in \cite{Lyutyy_energy} that thermal fluctuations do not modify 
the results but the frequencies where the enhancement is observed are too high for a medical treatment, 
i.e. for hyperthermia. Another example where the presence of uniaxial anisotropy was considered is 
Ref. \cite{Nandori}, where it was shown that in the low-frequency limit, an easy-axis anisotropy turned out 
to leave unchange (or slightly decrease) the calculated loss power if no transitions between the various 
precession modes (steady states) are taken into account which is the case for relatively moderate anisotropy. 
In summary, the alternating applied field is found to be slightly more favourable for low-frequencies which is 
suitable for hyperthermia and the small or moderate anisotropy does not change this picture.

Instead of trying to decide whether an alternating or a rotating external magnetic field is more suitable for 
magnetic nanoparticle hyperthermia, here we propose a new type of applied field, which is "simultaneously 
rotating and alternating". In particular, we demonstrate that a more efficient heat generation by magnetic 
nanoparticles is possible if the direction of the rotating external field changes periodically and the particles 
exhibit a moderate anisotropy. We show results of this new and successful attempt which enhances the loss 
energy drastically by means of abrupt changes in the applied magnetic field (sudden change of the direction 
of the rotating field after every circle). The change in the direction of the rotating applied field dislocates the 
magnetic moment of the nanoparticle out of its steady state, i.e. the precession mode. Since the steady state 
solution corresponds to minimal dissipation, a more effective heating can be achieved if the system is out of 
the steady state. We argue that the loss energy achievable by changing the direction of a rotating field is worth 
to study as a possible tool to enhance the heat in hyperthermia. The calculation has been done for a single, 
simple configuration, but the stunning result is worth checking experimentally. Indeed, we propose another 
new type of rotating applied field which has a periodically alternating direction, similarly to the case of "sudden 
change" studied in the present work but with a feature of being more suitable for experimental realisation.

The paper is organised as follows. In Sec~\ref{sec_llg}, we discuss the deterministic Landau-Lifshitz-Gilbert 
equation in case of uniaxial anisotropy with rotating applied field where overall units and parameters suitable 
for hyperthermia are also considered. Some known results on the specific loss power and loss energy obtained in 
the steady state solutions of the Landau-Lifshitz-Gilbert equation for oblate (positive anisotropy) particles is 
summaries briefly in Sec~\ref{sec_steady} and new results for prolate (negative anisotropy) particles are discussed.
New findings related to the loss energy per cycle out of the steady state are shown in Sec~\ref{sec_out}. 
In Sec~\ref{sec_new} we propose a novel type of applied field which is rotating with periodically changed direction. 
Proposal for experimental realization of the new applied field is discussed in Sec~\ref{sec_exp}. Finally, 
Sec~\ref{sec_sum} stands for the summary and a detailed discussion for the optimised set of parameters for 
magnetic particle hyperthermia is given.

\section{Landau-Lifshitz-Gilbert equation}
\label{sec_llg}
Out of the many phenomenological equations of motion for the relaxation of magnetization \cite{Berger} the 
Gilbert equation \cite{Gilbert} has proved to give the one of the most realistic description of the dynamics of 
single-domain magnetic particles at strong damping (Ref.~\cite{ferrohydro} represents another frequently used 
appoach). Such a particle, being too small to accommodate a domain wall, can be 
fully characterized with a single vector, its magnetic moment {\bf m}. An important feature of Larmor precession 
is that  $\vert {\bf m}\vert = m_S$ does not change under the influence of the external field, including the anisotropy field. 
Hence it is convenient to rewrite the equation of motion of the magnetization {\bf m} of a single-domain particle 
in terms of the unit vector ${\bf M} = {\bf m}/m_S$, $m_S$ being the saturation magnetic moment (e.g. 
$m_S \approx 10^5$ A/m for single crystal Fe$_3$O$_4$). Then the Gilbert equation reads as
\begin{equation}
\label{G}
\frac{\rm{d}}{\rm{d}t} {\bf M} = \gamma_0  {\bf M} \times  \left[\nabla_{\bf M}  V+ \mu_0 \eta\frac{\rm{d}}{\rm{d}t} {\bf M}\right],
\end{equation}
where  $\gamma_0 = 1.76 \times 10^{11}$ Am$^2$/Js  is the gyromagnetic ratio of the electron spin (with opposite 
sign), $\mu_0 = 4 \pi \times 10^{-7}$ Tm/A (or N/A$^2$) is the permeability of free space, $V$ is the potential energy 
and $\eta$ is the damping factor, both of them normalized for unit $M$. To describe the system, the potential energy 
must contain the Zeeman energy in the external applied magnetic field and the anisotropy energy \cite{Denisov2006}
\bea
V = -\mu_0 {\bf M} \cdot {\bf H}_{\rm{ext}} -\frac{\mu_0}{2} H_a M^2_z
\eea
wher $M_z$ is the z-component of the normalized magnetization vector and the external applied field is a rotating one
(perpendicular to the anisotropy field)
\bea
{\bf H}_{\rm{ext}}  = H_0 \, \, (\cos(\omega t), \sin(\omega t), 0),
\eea
with the angular frequency $\omega$. We define the vector ${\bf H}_{\rm{eff}}$, which contains the external applied 
magnetic field and the effect of the anisotropy of the magnetic particle
\bea
\label{H}
{\bf H}_{\rm{eff}} &=& -\frac{1}{\mu_0} \nabla_{\bf M}  V = - \frac{1}{\mu_0} (\partial_{M_x},\partial_{M_y},\partial_{M_z}) V \nn
&=& H_0 \, \, (\cos(\omega t), \sin(\omega t), \lambda_{\rm{eff}} M_z),
\eea
with $\lambda_{\rm{eff}} = H_a /H_0$. Clearly, if $\lambda_{\rm{eff}} > 0$ ($H_a > 0$), the anisotropy will turn the 
magnetization towards the z-axis, if  $\lambda_{\rm{eff}} < 0$ ($H_a < 0$), into the $xy$-plane. Giordano {\em et al.} 
\cite{Giordano} treated the shape anisotropy of the isotropic ellipsoidal particles using a different notation, where $L$ is 
the parameter giving the deviation from spherical symmetry. The link to our parameters is $H_a = (3L-1) m_S$. Although, 
our model does not specify the source of anisotropy, given the material used in hyperthermia, we have to deal with shape 
anisotropy.   

Equation \eq{H} implies a uniaxial anisotropy with a potential energy $-(\mu_0/2) H_0 \lambda_{\rm{eff}}  M^2_z$, which has its 
minima at $\vert M_z\vert = 1$ for $\lambda_{\rm{eff}} > 0$ and $M_z = 0$ for $\lambda_{\rm{eff}} < 0$. Two ellipsoids of 
revolution have the shapes with the geometry of this energy: for $\lambda_{\rm{eff}} > 0$ oblate (cigar-shape) and for 
$\lambda_{\rm{eff}} < 0$ prolate (lens shape). 

The Gilbert equation \eq{G} can be rewritten in such a way that it has a functional form similar to the Landau-Lifshitz 
equation \cite{LL}. This is called the Landau-Lifshitz-Gilbert (LLG) equation,
\begin{equation}
\label{LLG}
\frac{\rm{d}}{\rm{d}t} {\bf M} = -\gamma' [{\bf M \times H_{\rm{eff}}}] 
+ \alpha' [[{\bf M\times H_{\rm{eff}}]\times M}],
\end{equation}
where $\gamma' = \mu_0 \gamma_0 /(1+\alpha^2)$ and $\alpha' = \gamma' \alpha$ with the dimensionless 
damping factor $\alpha = \mu_0\gamma_0\eta m_S$. For example, $\alpha = 0.1$ was chosen in \cite{Lyutyy_energy}
and $\alpha = 0.3$ was used in \cite{Giordano}.
 
Writing Eq.~\eq{LLG} in polar coordinates ($M,\theta,\varphi$) allows to drop the constant (M), leaving but two equations:
\begin{align}
\label{rot_LLG}
\frac{\rm{d} \theta}{\rm{d}t} &=& 
\omega_L \sin\phi + \alpha_N \cos\theta \cos\phi 
- \alpha_N \lambda_{\rm{eff}} \sin\theta \cos\theta,
\nonumber \\
\frac{\rm{d} \phi}{\rm{d}t} &=& 
\omega_L \cos\phi \frac{\cos\theta}{\sin\theta} + \omega
- \alpha_N \frac{\sin\phi}{\sin\theta} 
- \omega_L \lambda_{\rm{eff}} \cos\theta
\end{align}
where $\omega_L = H_0 \gamma'$ and $\alpha_N = H_0 \alpha'$.
Note that \eq{rot_LLG} is written in a new coordinate system rotating with the applied field: the azimuthal angle 
($\varphi$) has been cut into the rotation ($\omega t$) and a measure ($\phi$) of the lag of {\bf  M} with respect to 
the rotation of the applied field: $\varphi = \omega t - \phi$.

Knowing that in the practice \cite{Johannsen2007} $H_0 \approx 18$ kA/m we find that $\omega_L$ is of the order 
of $10^9$ Hz. In hyperthermia the frequency of the applied field is advised to be chosen between  $10^5$ and 
$5\times10^5$ Hz, so that $\omega$ is four orders of magnitude below $\omega_L$. Furthermore, 
$\alpha_N = \alpha \, \omega_L $. For example, a set of parameters of \eq{rot_LLG} typical for hyperthermia
(with $\alpha = 0.1$ and $H_0 = 18$ kA/m),
\bea
\label{param}
\omega = 5 \times 10^5 \, \mr{Hz}, \,
\omega_L = 4 \times10^9 \, \mr{Hz}, \,
\alpha_N =  4 \times 10^8 \, \mr{Hz},
\eea
and the dimensionless anisotropy parameter $\lambda_{\mr{eff}}$ depends on the shape and geometry of the 
nanoparticle. The left sides of \eq{rot_LLG} being derivatives of angles with respect to time, the units of all terms 
in the equations must be s$^{-1}$. Their dimension can be taken of introducing a dimensionless "time" $\tilde t = t/t_0$
where $t_0 = 0.5 \times 10^{-10}$s is chosen in this work, thus, e.g. the dimensionless form of \eq{param} reads
\bea
\label{dimless_param}
\omega &\to&  \omega t_0 = 2.5 \times 10^{-5}, \nn
\omega_L &\to& \omega_L t_0  = 0.2, \nn
\alpha_N &\to& \alpha_N t_0  = 0.02.
\eea
Let us note, $t_0$ is chosen to be in the range of the attempt time $\tau_0$ used in Neel relaxation. Typical 
values for $\tau_0$ are between  $10^{-10}$  and $10^{-9}$ seconds.

\section{Steady state solution of the LLG equation}
\label{sec_steady}
\begin{figure}[ht] 
\begin{center} 
\epsfig{file=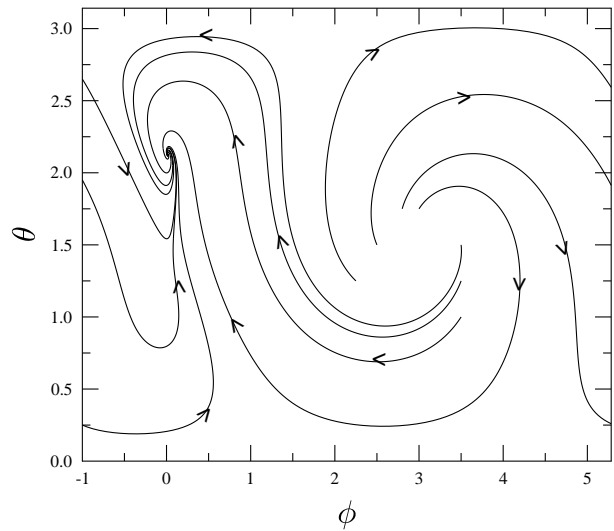,width=8.0 cm}
\caption{Orbit map in the rotating frame obtained by solving the LLG equation, 
below the critical value of anisotropy (there is a single attractive fixed point in 
the figure). The parameters are $\alpha_N = 0.1$, $\omega = - 0.01$, 
$\omega_L = 0.2$ and $\lambda_{\rm{eff}} = 1.1$. 
\label{fig1}
} 
\end{center}
\end{figure}

\begin{figure}[ht] 
\begin{center} 
\epsfig{file=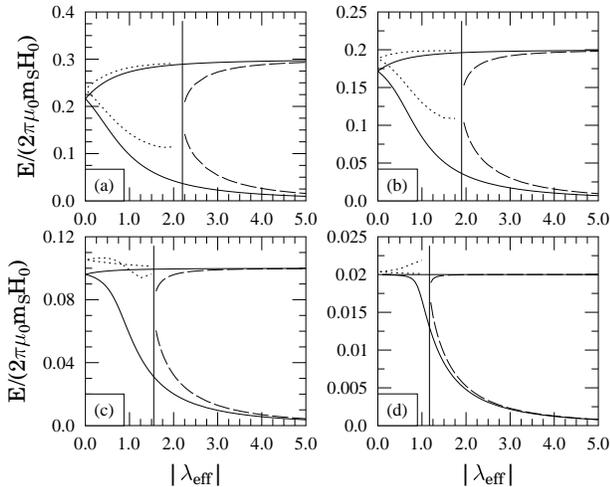,width=8.0 cm}
\caption{The loss energy per cycle as a function of the anisotropy parameter 
$\lambda_{\mr{eff}}$ is shown for various frequencies: (a) $\omega = 0.15$,
(b) $\omega = 0.1$, (c) $\omega = 0.05$ and (d) $\omega = 0.01$. The calculation
is based on the attractive fixed point solutions of \eq{rot_LLG} with $\alpha_N = 0.1$, 
$\omega_L = 0.2$.
The vertical lines indicate the critical value $\lambda_c$. Solid and dashed lines 
correspond to loss energies of steady state solutions. The upper lines (solid, dashed) 
always correspond to $\lambda_{\mr{eff}} <0$ while the lower lines (solid, dashed) are 
related to $\lambda_{\mr{eff}} >0$. Dotted lines represent the loss energy per cycle in 
case of rotating field where the direction is changed periodically. 
\label{fig2}
} 
\end{center}
\end{figure}

The solution of Eq.\eq{rot_LLG} is shown in \fig{fig1} for a particular set of (dimensionless) parameters given in 
the caption. As expected below the critical value of the anisotropy $\vert\lambda_{\mr{eff}}\vert < \lambda_c$,  
only a single attractive fixed point appears \cite{Nandori}, which corresponds to a steady state solution 
of the original LLG equation \eq{LLG} (single periodic precession mode),
\begin{eqnarray}
\label{steady_state}
M_{x}(t) &=& 
u_{x0} \cos(\omega t) - u_{y0} \sin(\omega t),
\nonumber \\
M_{y}(t) &=&
 u_{x0} \sin(\omega t) + u_{y0} \cos(\omega t),
\nonumber \\
M_{z}(t) &=& u_{z0}.
\end{eqnarray}
where $u_{x0}$ and $u_{y0}$ are determined by $\omega$, $\omega_L$, $\alpha_N$ 
and $\lambda_{\rm{eff}}$.
%
%
The loss energy is calculated using these attractive fixed point 
solutions in the formula for the energy dissipated in a single cycle,
\begin{eqnarray}
\label{def_loss}
E &=& \mu_0 m_S \int_{0}^{\frac{2\pi}{\omega}} \mr{d}t 
\left({\bf H}_{\rm{eff}} \cdot \frac{d{\bf M}}{dt} \right) \nn
&=& \mu_0 2\pi m_S H_0 (-u_{y0}),
\end{eqnarray}
(see also Eq.(12) in \cite{Nandori}) which has the following form in the low-frequency 
$\omega \ll \alpha_N$ and small anisotropy $\vert \lambda_{\mr{eff}} \vert \ll~1$ limits 
\bea
\label{loss_low-freq_small-aniso}
E \approx 2 \pi \mu_0 m_S H_0 \left[
\frac{\alpha_N \omega}{\omega_L^2 + \alpha_N^2}
- \frac{\alpha_N  \omega_L^2 \omega^3}{(\omega_L^2 + \alpha_N^2)^3}
(1+2 \lambda_{\mr{eff}}) \right].
\eea
It is clear that for positive (negative) 
$\lambda_{\rm{eff}}$ the energy per cycle is decreased (increased) by the anisotropy albeit for relatively 
large frequencies. Since $\lambda_{\mr{eff}}$ appears only at the next-to-leading terms, if one takes the 
low-frequency limit ($\omega \to 0$) the effect of anisotropy is irrelevant. This result has been discussed 
in \cite{Nandori} and can be seen in \fig{fig2}, which shows that for $\vert\lambda_{\mr{eff}}\vert < \lambda_c$, 
and for small frequencies, the loss energy per cycle becomes identical to that in the isotropic case, independently 
of the sign of $\lambda_{\rm{eff}}$. 

For large anisotropy, $\vert \lambda_{\mr{eff}}\vert > \lambda_c$, (if no transition between the various precession
modes is taken into account) the loss energy per cycle tends to zero for $\lambda_{\rm{eff}} > 0$ , and to a constant 
(almost identical to the isotropic value if $\omega \to 0$) for $\lambda_{\rm{eff}} < 0$. Similar observation holds for the 
loss energies in the steady states which exist only above the critical anisotropy, see the dashed lines in \fig{fig2}. For 
low frequencies, the dashed lines tend to zero for $\lambda_{\mr{eff}} > 0$ and to a constant for $\lambda_{\mr{eff}} <0$. 
As to the application in hyperthermia, these results confirm that particles of lens shape provide an enhanced 
loss energy, but at the allowed frequency the effect is not significant. 
%
%
It is important to note that frequencies ($\omega = 0.01 - 0.15$) shown in \fig{fig2} are too large for biomedical application.

\section{Loss energy per cycle out of the steady states}
\label{sec_out}
\begin{figure}[ht] 
\begin{center} 
\epsfig{file=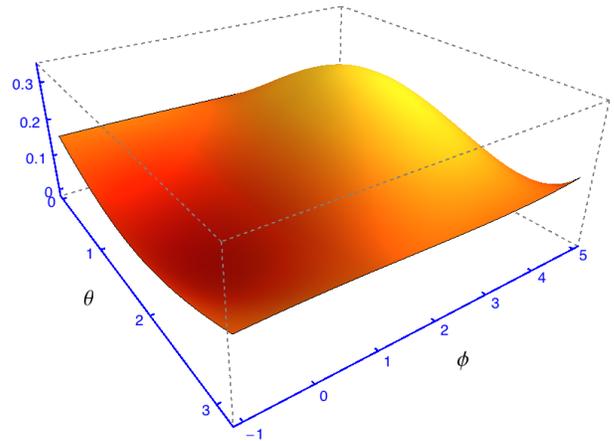,width=8.0 cm}
\caption{(Color online). The loss energy, $E/(2\pi \mu_0 m_S H_0)$, obtained in the first cycle (out of the steady 
states) for small anisotropy ($\lambda_{\mr{eff}} = 0.05$) as a function of the initial conditions on 
the ($\theta,\phi$) plane. 
\label{fig3}
} 
\end{center}
\end{figure}

Up to now, we studied the loss energy per cycle calculated at the steady state solution of the
LLG equation (i.e. at the attractive fixed point of \fig{fig1}). However, it is a relevant question 
to address whether one can find a larger loss energy in a single cycle if it is calculated out of 
the steady states. Indeed, in the 3D-plot of \fig{fig3} we show the loss energy obtained in the
first cycle of the external field as a function of various starting points on the ($\theta,\phi$) plane. 
%
%
Note that in the low-frequency limit (suitable for hyperthermia), the solution of the 
LLG equation always tends  to the attractive fixed point very rapidly, i.e. it reaches the 
fixed point with one percent accuracy in a quarter of a full circle. In \fig{fig3} the "well" 
corresponds to the attractive fixed point which produces us the lowest loss energy per 
cycle while the "hill" where the loss energy is the maximum is related to initial conditions 
taken at the repulsive fixed point. Though, the results displayed in \fig{fig3} are obtained 
numerically, an approximative analytic derivation is possible for the isotropic case 
($\lambda_{\mr{eff}} = 0$). Indeed, an approximate solution to the LLG equation is given
by Eq.(29) of \cite{Chatel} which reads
\begin{eqnarray}
\label{rot_sol_relax_approx_a}
M_{\xi}(t) &=&  
M_{\xi 0}^{\mr{spec}} \cos(\Omega t) - M_{\eta 0}^{\mr{spec}} \sin(\Omega t) \nn
&+& (M_{\xi 0} - M_{\xi 0}^{\mr{spec}}) \, \, e^{-\frac{\alpha_N t}{\sqrt{2}}},
\nonumber \\
M_{\eta}(t) &=& 
M_{\xi 0}^{\mr{spec}} \sin(\Omega t) + M_{\eta 0}^{\mr{spec}} \cos(\Omega t) \nn
&+& (M_{\eta 0} - M_{\eta 0}^{\mr{spec}}) \, \, e^{-\frac{\alpha_N t}{\sqrt{2}}},
\nonumber \\
M_{\zeta}(t) &=& 
\sqrt{1-[M_{\xi}(t)]^2 -[M_{\eta}(t)]^2},
\end{eqnarray}
where $(M_{\xi}, M_{\eta}, M_{\zeta})$ represents the magnetization vector in a particular rotated 
coordinate system \cite{Chatel} and depends on the initial values of the cartesian coordinates 
$M_{\xi 0}$ and $M_{\eta 0}$ linearly. ($M_{\xi 0}^{\mr{spec}}$, $M_{\eta 0}^{\mr{spec}}$ depend 
on the parameters $\omega, \omega_L, \alpha_N$ defined by Eq.(28) of \cite{Chatel} and 
$\Omega=\sqrt{\omega^2 + \omega_L^2}$). According to \eq{def_loss}, the loss energy per cycle 
has a linear dependence on the initial cartesian coordinates $M_{\xi 0}$ and $M_{\eta 0}$ as well. 
Thus, rewriting them by polar coordinates 
\bea
M_{\xi 0} \propto \sin(\theta) \cos(\phi), \hskip 0.5cm
M_{\eta 0} \propto \sin(\theta) \sin(\phi)
\eea
one finds
\bea
E/(\mr{1st \,\, cycle}) = A \sin(\theta) \cos(\phi) + B \sin(\theta) \sin(\phi)
\eea
where $A,B$ constants depend on the parameters $\omega,\omega_L, \alpha_N$ which is in 
agreement to the numerical result of \fig{fig3} obtained for small anisotropy. Finally, let us note that
the numerical results of \fig{fig3} are related to the fact that the steady state solution corresponds to 
minimal dissipation, so, a more effective heating can be achieved if the system is out of the steady state. 
This preliminary result does not define a new method to enhance the loss energy but the difference between 
the maximum and minimum on \fig{fig3} being two orders of magnitude larger than what can be achieved 
in the steady state, is encouraging for further research on non-steady states.

\section{Rotating field with periodically changed direction}
\label{sec_new}
As a new idea let us consider the loss energy per cycle in case of a rotating applied magnetic field where 
the direction of rotation is changed periodically, i.e., after every full cycle the sign of $\omega$ is switched 
in \eq{H}. By changing the direction, the position of the steady state solution is changed, i.e. for $\omega >0$ 
($\omega < 0$) it is below (above) the equator of the orbit map. Thus, the magnetization vector always tends 
from a disappearing steady state to an arising one. Note that we do not calculate this effect above $\lambda_c$ 
because in that case more than one stable steady state solutions appear. The outcome is plotted in \fig{fig2}, 
with dotted lines. 

For high frequencies ($\omega = 0.15$ and $\omega = 0.1$), irrelevant to the sign of $\lambda_{\mr{eff}}$, 
there is an increase in the loss energy, tending to constant values close to $\lambda_c$. The effect is larger 
for $\lambda_{\mr{eff}}>0$, because in this case the positions of the steady state solutions (for $\omega <0$ 
and $\omega >0$) are far from each other. 

\begin{figure}[ht] 
\begin{center} 
\epsfig{file=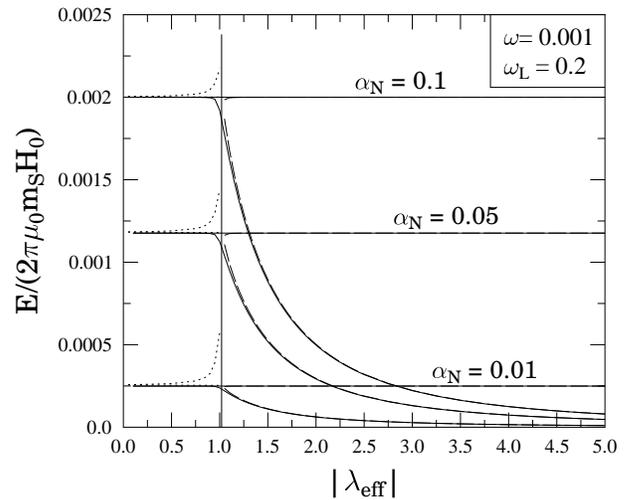,width=8.0 cm}
\caption{The loss energy per cycle as a function of the anisotropy parameter 
$\lambda_{\mr{eff}}$ is shown for various values of $\alpha_N$ (for fixed $\omega$
and $\omega_L$).
\label{fig4}
}
\end{center}
\end{figure}

For low frequencies ($\omega = 0.001$) the situation completely changes as the loss energy for $\lambda_{\rm eff} > 0$ 
becomes larger than any other calculated values, including those found for $\lambda_{\rm eff} < 0$. For example, for
lower values of the damping, i.e. $\alpha_N = 0.05$ or $\alpha_N = 0.01$ the loss energy monotonously increases 
with $\lambda_{\mr{eff}}$, reaching a value that is about 15\% or 100\% larger as compared to the isotropic case 
near (below) $\lambda_c$, see \fig{fig4}. Therefore, a more efficient heat generation is possible for physical values of 
the damping (i.e. for $\alpha = 0.1 - 0.3$ where $\alpha_N = \alpha \omega_L \approx 0.02 - 0.06$), near the 
critical anisotropy $\lambda_{\mr{eff}} = \lambda_c \approx 1$ if the direction of the rotating applied field is 
changed periodically.
%
%
In summary, the reason for the peaks of the dissipated energy at $\vert \lambda_{\mr{eff}} \vert=1$ in \fig{fig4} 
(dotted lines) is twofold: (i) it was shown that a more effective heating can be achieved if the system is out of the steady 
state which is achieved by the new type of applied field (dotted lines) which is simultaneously rotating and alternating, 
(ii) this enhancement effect becomes larger at $\vert \lambda_{\mr{eff}} \vert=1$ where the system is close to the critical 
anisotropy $\vert \lambda_{c} \vert \sim 1$, where the dislocating effect caused by the change in the direction of the 
rotation is the largest.

\section{Proposal for experimental realization }
\label{sec_exp}
Finally, let us propose a new type of "rotating" applied field which has a periodically alternating direction,
similarly to the case carefully analyzed above but with a feature of being more suitable for experimental 
realization. Nonetheless, a "sharp change" in the direction of the rotating field which is studied previously is 
also feasible in practice. Our proposal for the applied field is 
\begin{equation}
\label{new}
{\bf H}_{\rm{eff}} = H_0 \, \, (\cos(2\omega t), \sin(\omega t), \lambda_{\rm{eff}} M_z),
\end{equation}
where the $2 \omega$ of the cosine naturally provides us with the required change in the direction in every
half cycle. Well separated and oriented anisotropic nanoparticles in an aerogel matrix can serve as a good 
experimental setup to realize the enhanced loss energy in case of an applied field as in Eq. \eq{new}.
The practical advantage of \eq{new} is that it requires the use of two frequencies $\omega$ and $2\omega$ 
and no higher harmonics or no rapid switchings are needed. Of course, one has to choose the frequency 
$\omega$ in a way that $2\omega$ should be in the range suitable for hyperthermia.

\section{Summary}
\label{sec_sum}
In this work we showed that a more efficient heat generation by magnetic nanoparticles is possible if the 
direction of the rotating external field changes periodically. We would like to emphasise that the heating 
efficiency of the new type of applied field is larger than those of the rotating and alternating ones which are 
almost identical in the limit of small frequency, see for example \cite{Ahsen2010, Raikher}. Based on this finding, 
we make a proposal to apply a magnetic field regularly stopping its rotation, which should increase the loss energy. 
Conditions used in the proposed work towards an enhance of heating efficiency are the followings.
\\
{\bf Applied field.--} A new type of "rotating" applied field is proposed here which has a periodically alternating 
direction either using the form \eq{H} where $\omega$ is assumed to change sign in every full circle or 
using \eq{new} where the $2 \omega$ of the cosine naturally provides us with the required change in the direction 
in every half cycle. The frequency is in the range of hyperthermia, i.e. $\omega = 1-5 \times 10^5$ Hz and the
amplitude of the applied filed is $H_0 \approx 0.2 \times 10^5$A/m.
Note that \eq{new} produces a change in the rotation in every half circle compared to the sudden change where 
the direction of the rotation changes in every full circle. We argued that in the low-frequency limit, the solution of the 
LLG equation always tends to the attractive fixed point very rapidly, i.e. it reaches the fixed point with one percent 
accuracy in a quarter of a full circle. Thus, it is expected that the proposed external field \eq{new} possesses the 
same properties found for the case of sudden change.
\\
{\bf Anisotropy.--} The nanoparticles are assumed to be oblate ellipsoids where the shape anisotropy is moderate, 
i.e. $\lambda_{\mr{eff}} = H_a/H_0 \approx 1$ (which implies $L \approx1.2/3 = 0.4$ according to Ref. \cite{Giordano}
for $H_a = 0.2 \times 10^5$A/m and $M_s = 10^5$A/m). In this case the nanoparticle has a single precession mode. 
However, for larger anisotropy, i.e. $\lambda_{c} < \lambda_{\mr{eff}}$ which has been discussed in \cite{Lyutyy_energy} 
(where $\tilde h = \lambda_{\mr{eff}}^{-1}$), more than one precession modes appear. In Ref. \onlinecite{Lyutyy_energy} an 
enhancement of the heating efficiency was found near the boundary of these regimes of forced precession. The largest 
increase is observed between the periodic and quasi periodic regimes, however, it exists only for large values of the 
reduced frequency used in \cite{Lyutyy_energy} which is not allowed in hyperthermia. Nevertheless, it is expected that 
in case of an applied field, Eq.~\eq{new} which can be considered as a rotating field (with alternating direction) where the 
magnitude changes slightly, the effect of the increased upward heating becomes even stronger due to the transition between 
the two periodic modes. Finally, let us note that regarding the anisotropy, one has to take into account the structure formation 
and synchronization of interacting magnetic nanoparticles which have been studied e.g., in \cite{structure} for rotating fields. 
\\
{\bf Orientation.--} The anisotropy field (shape anisotropy caused by oblate particles) is assumed to be perpendicular
to the plane where the applied field rotates. In practice, this special situation can be achieved by switching on a strong 
static field which orients all the nanoparticles and than the rotating field applied while the static field switched off. For an
appropriate size of the nanoparticle (with an average diameter $d=14$nm \cite{Vallejo}), the orientation of the major part 
of the nanoparticles remains unchanged (for long enough time) and only the magnetic moment has a dynamics under the 
rotating field. Number of parameters (for example biomedical coating) can be tuned in order to design the nanoparticles 
(see e.g. \cite{design}) for biomedical application incorporating the requirement for particle size and shape-anisotropy.
Furthermore, in \cite{Lyutyy_energy} it was argued that if the average diameter is chosen to be in the range $(d_{\mr{min}},d_{\mr{max}})$ 
than the nanoparticles have single-domains and one finds no restriction on the applied frequency and their dynamics is
almost deterministic. The boundary values, $d_{\mr{min}}$ and $d_{\mr{max}}$ depend on the parameters such as the 
anisotropy factor: e.g., in case of relatively large anisotropy ($\tilde h = 0.1$) one finds $d_{\mr{min}} = 13.7$nm. Thus,
our choice $d=14$nm indicates that the description of the dynamics of the nanoparticle magnetic moments by the 
deterministic LLG equation is well justified \cite{Lyutyy_energy} and it is expected that they retain their orientation 
for long enough time \cite{Vallejo}. 
\\
{\bf Damping.--} The dimensionless damping constant is chosen to be in the range $\alpha = 0.1 - 0.3$ which was 
used in the works \cite{Giordano,Lyutyy_energy}. Accordingly, the enhancement effect shown in this work is 100\% - 15\%
(in case of a smaller damping the enhancement is larger).

\section*{Acknowledgement}
This work was supported by the project T\'AMOP-4.2.2.A-11/1/KONV-2012-0036 co-financed by the European Union and the 
European Social Fund. Partial support is acknowledged to the János Bolyai Research Scholarship of the Hungarian Academy 
of Sciences and to the National Research, Development and Innovation Office of Hungary under project No. K115575.


\begin{thebibliography}{99}

\bibitem{ferrofluid}
C. Scherer, H.G. Matuttis, Phys. Rev. E {\bf 63} 011504 (2000);
J. Embs, H. W. M\"uller, C. Wagner, K. Knorr, and M. L\"ucke, Phys. Rev. E {\bf 61}, R2196 (2000);
B. U. Felderhof, Phys. Rev. E {\bf 62}, 3848 (2000);
M.I. Shliomis, Phys. Rev. E {\bf 64}, 063501 (2001);
B. U. Felderhof, Phys. Rev. E {\bf 64}, 063502 (2001);
H.W. M\"uller and M. Liu, Phys. Rev. E {\bf 64}, 061405 (2001);
J. P. Embs, S. May, C. Wagner, A. V. Kityk, A. Leschhorn, and M. L\"ucke, Phys. Rev. E {\bf 73}, 036302 (2006);
A. Leschhorn, M. L\"ucke, C. Hoffmann and S. Altmeyer, Phys. Rev. E {\bf 79}, 036308 (2009).

\bibitem{ferrohydro}
M.I. Shliomis, Zh. Eksp. Teor. Fiz. {\bf 61}, 2411 (1971) [Sov. Phys. JETP {\bf 34}, 1291 (1972)];
M.I. Shliomis, Usp. Fiz. Nauk {\bf 112}, 427 (1974) [Sov. Phys. Usp. {\bf 17}, 153 (1974)];.
R.E. Rosensweig, Ferrohydrodynamics, Cambridge University Press, Cambridge, (1985);
M.I. Shliomis, Phys. Rev. E {\bf 64}, 060501(2001);
 
\bibitem{biomedical}
Q. A. Pankhurst, J. Connolly, S. K. Jones, and J. Dobson, J. Phys. D: Appl. Phys. {\bf 36}, R167 (2003);
M. Ferrari, Nat. Rev. Cancer {\bf 5}, 161 (2005);
R. Hergt, S. Dutz, R. Muller, M. Zeisberger, J. Phys.: Condens. Matter {\bf 18} S2919 (2006);
R. Hergt, S. Dutz, J. Magn. Magn. Mater. {\bf 311}, 187 (2007);
S. Laurent, D. Forge, M. Port, A. Roch, C. Robic, L. Vander Elst, and R. N. Muller, Chem. Rev. {\bf 108}, 2064 (2008);
J. D. Alper, Thesis (Ph.D.), Massachusetts Institute of Technology (2010); 
Quian Wang and Jing Liu, Fundamental Biomedical Technologies {\bf 5}, 567-598 (2011), 
Spinger Science+Business Media: {\em Intercellular Delivery} (ed. A. Prokop);
A. L. E. Rast, Thesis (Ph.D.) University of Alabama, Birmingham (2011);
D. E. Bordelon, C. Cornejo, C. Gr\"uttner, F. Westphal, T. L. DeWeese, R. Ivkov, J. Appl. Phys. {\bf 109}, 124904 (2011);
M. V. Petrova, {\em et al.}, Applied Magnetic Resonance {\bf 41}, 525 (2011);
A. Arakaki, K. Shibata, T. Mogi, M. Hosokawa, K. Hatakeyama, H. Gomyo, T. Taguchi, H. Wake, T. Tanaami, 
T. Matsunaga and T. Tanaka, Polymer Journal {\bf 44}, 672 (2012).
S Bucak, B Yavuzt\"urk, A D Sezer, 
Recent Advances in Novel Drug Carrier Systems, Ali Demir Sezer (Ed.), ISBN: 978-953-51-0810-8, InTech. (2012)

\bibitem{Stauffer}
P. R. Stauffer Int. J. Hyperthermia {\bf 21} (2005) 731.

\bibitem{Johannsen}
M. Johannsen, U. Gneveckow, K. Taymoorian, C.H. Cho, B. Thiesen, R. Scholz, 
N. Wald�fner, S.A. Loening, P. Wust, A. Jordan Acta Urol. Esp. {\bf 31} (2007) 660.

\bibitem{Bellizzi}
G. Bellizzi and O.M. Bucci, Int. J. Hyperthermia {\bf 26} (2010) 389-403.

\bibitem{Bertotti} 
Giorgio Bertotti, Claudio Serpico, and Isaak D. Mayergoyz,
Phys. Rev. Lett. {\bf 86} (2001) 724.

\bibitem{Denisov2006}
S. I. Denisov, T. V. Lyutyy, P. H\"anggi, Phys. Rev. Lett. {\bf 97}, 227202 (2006).
S. I. Denisov, T. V. Lyutyy, P. H\"anggi, and K. N. Trohidou, Phys. Rev. B {\bf 74}, 104406 (2006).

\bibitem{Chatel}
P. F. de Ch\^atel, I. N\'andori, J. Hakl, S. M\'esz\'aros and K. Vad,
J. Phys.: Condens. Matter {\bf 21} (2009) 124202.

\bibitem{Cantillon}
P. Cantillon-Murphy, L.L. Wald, E. Adalsteinsson, M. Zahn, JMMM {\bf 322}, 727 (2010).

\bibitem{Ahsen2010}
O. O. Ahsen, U. Yilmaz, M. D. Aksoy, G. Ertas, E. Atalar, JMMM {\bf 322}, 3053 (2010).

\bibitem{Denisov_thermal}
S. I. Denisov, T. V. Lyutyy, C. Binns, P. H\"anggi, JMMM {\bf 322}, 1360 (2010);
S. I. Denisov, A. Yu. Polyakov, and T. V. Lyutyy, Phys. Rev. B {\bf 84}, 174410 (2011).

\bibitem{Raikher}
Yu. L. Raikher and V. I. Stepanov, Phys. Rev. E {\bf 83}, 021401 (2011).

\bibitem{Nandori} 
I. N\'andori, J. R\'acz, Phys. Rev. E {\bf 86} (2012) 061404.

\bibitem{Lyutyy} 
T. V. Lyutyy, S. I. Denisov, A. Yu. Polyakov, C. Binns,
INTERNATIONAL CONFERENCE NANOMATERIALS: APPLICATIONS AND PROPERTIES {\bf 1} (2012) 04MFPN16.

\bibitem{Lyutyy_energy}
T. V. Lyutyy, S. I. Denisov, A. Yu. Peletskyi and C. Binns, Phys. Rev. B. {\bf 91}, 054425 (2015). 

\bibitem{Chen}
Shih-Wei Chen, Jr-Jie Lai, Chen-Li Chiang and Cheng-Lung Chen, 
REVIEW OF SCIENTIFIC INSTRUMENTS {\bf 83}, 064701 (2012).

\bibitem{stochastic_llg_lin}
H. El Mrabti, S. V. Titov, P.M. D\'ejardin, Y. P. Kalmykov, J. Appl. Phys. {\bf 110}, 023901 (2011);
H. El Mrabti, P.M. D\'ejardin, S. V. Titov, Y. P. Kalmykov, Phys. Rev. B {\bf 85}, 094425 (2012).

\bibitem{Berger}
R. Berger, J/C Bissey and J. Kliava, J. Phys.: Condens Matter {\bf 12} (2000) 9347.

\bibitem{Gilbert}
T. L. Gilbert, Phys. Rev. {\bf 100} (1955) 1243.

\bibitem{LL}
L. Landau and E. Lifshitz, Phys. Z. Sowjetunion {\bf 8}, 153 (1935).

\bibitem{Giordano}
S. Giordano, Y. Dusch, N. Tiercelin, P. Pernod, V. Preobrazhensky, Eur. Phys. J. B {\bf 86}, 249 (2013). 

\bibitem{Johannsen2007}
M. Johannsen, U. Gneveckow, K.Taymoorian, C.H. Cho, B. Thiesen, R. Scholz, N. Wald�fner, S.A. Loening, P. Wust and A. Jordan, 
Actas Urol. Esp. {\bf 31} (2007) 660.

\bibitem{Fannin}
P. C. Fannin, I. Malaescue and C. N. Marin, JMMM {\bf 289}, 162 (2005). 

\bibitem{Vallejo}
G. Vallejo-Fernandez, O. Whear, A. G. Roca, S. Hussain, J. Timmis, V. Patel and K. O'Grady
J. Phys. D: Appl. Phys. {\bf 46}, 312001 (2013).

\bibitem{design}
A. Hajdu, {\em et al.}, Colloids and Surfaces B: Biointerfaces {\bf 94}, 242 (2012);
I. Y. Toth, {\em et al.}, Langmuir {\bf 28}, 16638 (2012);
M. Szekeres, {\em et al.}, Int. J. Mol. Sci. {\bf 14}, 14550 (2013);
I. Y. Toth, {\em et al.}, Langmuir {\bf 30}, 15451 (2014);
M. Szekeres, {\em et al.}, Nanomedicine and Nanotechnology, {\bf 6}, 1000252 (2015);
I. Y. Toth, {\em et al.}, JMMM {\bf 380}, 168 (2015).

\bibitem{structure}
J. E. Martin, R. A. Andreson, C. P. Tigges, J. Chem. Phys. {\bf 108}, 7887 (1998)
J. E. Martin, R. A. Andreson, C. P. Tigges, J. Chem. Phys. {\bf 110}, 4854 (1999);
S. J\"ager and S. H. L. Klapp, Soft Matter {\bf 7}, 6606 (2011).

\end{thebibliography}
\end{document}